\documentclass[sort&compress,preprint,10pt]{elsarticle}
\usepackage{graphicx}
\usepackage{amsmath}
\usepackage{amssymb}
\usepackage{epsfig}
\usepackage{subfigure}
\usepackage{wrapfig}
\usepackage{fancyvrb}

\newcommand{\be}{\begin{equation}}
\newcommand{\ee}{\end{equation}}

\journal{Computer Physics Communications}

\begin{document}

\begin{frontmatter}
\title{\vspace*{2cm}Multidimensional spline integration of scattered data
}

\author[rvt,rvt2]{Gergely Endr\H{o}di}
\ead{endrodi@general.elte.hu}
\address[rvt]{Institute for Theoretical Physics, E\"otv\"os University, H-1117 Budapest, Hungary.}
\address[rvt2]{Institute for Theoretical Physics, Universit\"at Regensburg, D-93040 Regensburg, Germany.}

\begin{abstract}
We introduce a numerical method for reconstructing a multidimensional surface using the gradient of the surface measured at some values of the coordinates. The method consists of defining a multidimensional spline function and minimizing the deviation between its derivatives and the measured gradient. Unlike a multidimensional integration along some path, the present method results in a continuous, smooth surface, furthermore, it also applies to input data that are non-equidistant and not aligned on a rectangular grid. Function values, first and second derivatives and integrals are easy to calculate. The proper estimation of the statistical and systematical errors is also incorporated in the method.
\end{abstract}


\end{frontmatter}

\section{Introduction}

Experiments usually result in a discrete set of datapoints $F_i$ measured at a discrete set of coordinates $\mathbf{q}_i$.
It is in general useful to interpolate this discrete set to obtain a smooth surface $S(\mathbf{q})$ as a function of the coordinates. A more complicated situation is when derivatives $\partial F/\partial q_i$ are measured, and the surface $S(\mathbf{q})$ which fits on this grid of gradients the best is sought for.

The latter case is commonly encountered in statistical physical problems, e.g. in lattice field theory, where using standard Monte-Carlo methods one is not able to measure the (logarithm of the) partition function $\log \mathcal{Z}$ itself, but only its derivatives with respect to the parameters of the theory. These parameters play the role of the above coordinates $\mathbf{q}$, and the free energy $-\log \mathcal{Z}$ as a function of the parameters is what one is after. A multidimensional integration is usually carried out in such situations, which, however, can only be applied when the measurements reside on a rectangular grid. Furthermore, the result will also depend on the integration path and is only guaranteed to be smooth along the path itself.

The smooth interpolation throughout the whole grid can be obtained by e.g. a multidimensional spline function. For a detailed discussion of splines and their application see e.g. Ref.~\cite{Ahlberg}, Ref.~\cite{deBoor} or Ref.~\cite{Schumaker}. Methods to determine a two dimensional spline surface upon a rectangular grid given the values of the surface at the grid points have been known for a long time (see e.g. Ref.~\cite{Christie:1978sa}), along with algorithms for nonrectangular input datasets (see e.g. Ref.~\cite{Friedman:1980sn,Franke:1982}), using once again the {\it values} of the surface at given coordinates. Since then spline approximation has received quite a lot of attention and the mathematical basis was studied in detail. For recent reviews on these approaches see e.g. Refs.~\cite{Nurnberger:2000, Goodman:2006}.

Further methods for multivariate approximation have also been developed, e.g. using rational basis functions, B-splines, tensor product splines, Powell-Sabin splines, triangulations, genetic algorithms or modified spline techniques. Moreover, based on such methods, various algorithms for spline fitting are also present in the literature, see Refs.~\cite{Wachspress:1971, Wang:1993, Park:2010, Quak:1990, Fong:1993, Davydov:2006, Goshtasby:1993, Franke:1999, Constantini:1999, Dierckx:1981, Juttler:1997, Juttler:1999, Dierckx:1991, Dierckx:1994, Markus:1997, Yoshimoto:2003, Nunhez:2004}. Techniques for handling discontinuities (e.g. Ref.~\cite{Silanes:2001}) and imposing constraints on the fitted function can also be found (e.g. Refs.~\cite{Franke:1983a, Rogers:1989}).

In this paper we present a method which determines the smooth surface $S(\mathbf{q})$ using the {\it gradient} of the surface measured at some scattered values of the coordinates, and thus corresponds to a {\it multidimensional integration scheme}. The algorithm includes an introduction of a set of nodepoints, upon which the multidimensional spline is determined. This determination is linear and thus straightforward to compute. By a systematical variation of the number and position of the nodepoints the method is adapted to the particular problem, which is highly important for interpolation in more than one dimensions, as pointed out in Ref.~\cite{Friedman:1980sn}. Other interpolation schemes for which the parametrization is adapted to the data (like the rational basis function approach) are also well suited for the present problem and may be applied in this scope.

The paper is structured as follows.
In section~\ref{sec:splinedef} we remind the reader how a spline function in an arbitrary number of dimensions $D$ is defined and then in section~\ref{sec:splinefit} we show how the fit to the measurements is carried out. Since in practice $D>2$ is seldom necessary, in order not to complicate the notation we present the method in detail for the case of two dimensions. Nevertheless, $D>2$ is also straightforward to implement. Then we present the algorithm for the systematical placement of the nodepoints, and finally show several results on mock data.

\section{Spline definition}
\label{sec:splinedef}

In two dimensions a cubic spline is defined upon a grid $\{x_k, y_l\}$ with $0\le k <K$, $0\le l<L$. The spline surface is unambiguously determined by the values that it takes at the nodepoints $f_{k,l}=S(x_k,y_l)$ (and the boundary conditions, which we specify to be ``natural'', see later). A grid square $[x_k,x_{k+1}]\times[y_l,y_{l+1}]$ will be shortly referred to as $\{k,l\}$. The spline function itself is compactly written as\footnote{Note that in this formulation the cubic spline contains terms like $x^3y^3$, contrary to other definitions where a bicubic spline only has terms $x^{i}y^{j}$ with $0\le i+j\le3$.}
\be
S(x,y)=\sum\limits_{i=0}^{3}\sum\limits_{j=0}^{3} C_{i,j}^{k,l} t_{(k)}^i u_{(l)}^j,\quad \textmd{if } (x,y)\in\{k,l\}
\label{eq:sdef}
\ee
where $C_{i,j}^{k,l}$ are the spline coefficients and $t_{(k)}$ and $u_{(l)}$ the dimensionless coordinates:
\be
t_{(k)} \equiv \frac{x-x_{k}}{w_{k}^{(x)}},\quad\quad u_{(l)} \equiv \frac{y-y_{l}}{w_{l}^{(y)}}
\label{eq:t2}
\ee
with the widths
\be
w_{k}^{(x)} \equiv x_{k+1}-x_{k},\quad\quad w_l^{(y)} \equiv y_{l+1}-y_{l}
\ee
The linear equations that determine the spline parameters $f_{k,l}$ from the coefficients for the $x$ and $y$ directions (summarized in matrix form as $X$ and $Y$, respectively) are independent from each other. Inverting these equations one obtains for the coefficients
\be
C_{i,j}^{k,l} = \sum\limits_{n_1=0}^{K-1}\sum\limits_{n_2=0}^{L-1} (Y^{-1})_{4l+j}^{n_2} (X^{-1})_{4k+i}^{n_1} f_{n_1,n_2}
\label{eq:c2}
\ee
Here the matrix $X$ ($Y$) only depends on the number $K$ ($L$) of grid points in the $x$ ($y$) direction and the widths $w_k^{(x)}$ ($w_l^{(y)}$). The code for generating the matrix $X$, together with a sample output is shown in~\ref{sec:app1}.
Note that $X$ is a matrix of size $4(K-1)\times 4(K-1)$, but in order to obtain the coefficients we only need the first $K$ rows of its inverse -- or, even better, the effect of these rows on the vector $f$.

Similarly, in arbitrary dimensions $D$ the different directions decouple and thus~(\ref{eq:c2}) is easily generalized to the case of more dimensions. In the following we introduce the fitting method in two dimensions; the generalization for higher dimensions is also straightforward to implement.

We note here that a compact representation of splines can also be achieved with the help of B-splines (for an overview see Refs.~\cite{Prautzsch, Hoschek, Piegl}). In this approach it is not necessary to prescribe particular boundary conditions, as for the spline function above (we use ``natural'' boundary conditions in the definition of $X$). 

\section{Spline fitting}
\label{sec:splinefit}

For any value of the parameters $f_{k,l}$ the coefficients $C_{i,j}^{k,l}$ -- i.e. the whole spline surface $S(x,y)$ -- are known unambiguously. This way we consistently parameterized the spline surface. Now we want to set the spline parameters $f_{k,l}$ such, that derivatives of the spline surface are as close to some previously measured values as possible. Note that this way the function $S(x,y)$ will be undetermined upto an overall constant, since the translation $S(x,y)\to S(x,y)+A$ does not influence the derivatives. This symmetry will be taken into account in the following.

Let us consider $N$ number of points and say that at each point $(q^{(x)}_m,q^{(y)}_m)$ we measured the derivative\footnote{By measurement we refer to e.g. the lattice determination of the derivatives of $\log \mathcal{Z}$ at a given value of some bare parameters.} in the $x$ and $y$ directions: $D^{(x)}_m$ and $D^{(y)}_m$ with errors $\Delta D^{(x)}_m$ and $\Delta D^{(y)}_m$ ($m=0\ldots N-1$). Being ``close'' can be quantified by minimizing
\be
\chi^2(f_{k,l})=\sum\limits_{m=0}^{N-1}\left[ \left( \frac{\frac{\partial S}{\partial x} - D^{(x)}_m}{\Delta D^{(x)}_m}\right)^2 + \left(\frac{\frac{\partial S}{\partial y} - D^{(y)}_m}{\Delta D^{(y)}_m}\right)^2 \right ]
\label{eq:chisqr}
\ee
Note that the above $\chi^2$ represents a situation where the input data for the derivatives $D^{(x)}$ and $D^{(y)}$ are uncorrelated. However, this is usually not the case and one has to take into account the correlation between the measurements. Generalization of the method to include this correlation is presented in~\ref{sec:app2}.

Since $S$ and thus $\partial S/\partial x$ and $\partial S/\partial y$ are linear in $f_{k,l}$, this function has a quadratic dependence on the parameters $f_{k,l}$. This enables us to search for the minimum of $\chi^2(f_{k,l})$
\be
\frac{\partial \chi^2}{\partial f_{n_1,n_2}} = 0
\ee
by solving a system of linear equations
\be
M_{n_1,n_2}^{k,l} f_{k,l} = V_{n_1,n_2}, \quad n_1=0\ldots K-1,\;\; n_2=0\ldots L-1
\label{eq:sys}
\ee

Due to the above mentioned translational invariance of the solution, this system of equations is underdetermined and thus the inverse of $M$ does not exist. This can also be seen by checking that $M$ has a zero eigenvalue, corresponding to the eigenvector $(1,1,1,\ldots)$, or, in other words, each row of $M$ adds up to zero. Physically this means that one can set e.g. the first element of $f$ to zero, i.e. leave the first column of $M$. To obtain an invertible matrix one now has to drop one of its rows, for example the first\footnote{It can be checked that after eliminating the first column, any row of the matrix $M$ can be reproduced by a linear combination of the other rows, i.e. it is indeed correct to drop an arbitrary row.}. This way one arrives at a matrix $M'$ of size $(KL-1)\times (KL-1)$. In the same manner we define $V'$ to be the vector composed from the last $KL-1$ elements of $V$. One can then complement the solution $(M')^{-1}V'$ with a zero in the first element to obtain the final result which satisfies $f_{0,0}=0$.

We remark here that in some cases it can happen that the matrix $M$ is not invertible. An obvious example is when there exists a grid square $\{k,l\}$ in which no measurements reside. The spline function is then ambiguous in this grid square and thus $M$ necessarily has a rank smaller than $KL-1$.
Furthermore, an unlucky choice of the coordinates of the measurements $q^{(x)}$,$q^{(y)}$ may further reduce the rank of the matrix $M$, but this is very unlikely. By a systematic replacement of the nodepoints most of such situations may be filtered out, see section~\ref{sec:stable}.

In order to explicitly write the elements of the matrix $M$ and the vector $V$ let us analyze the dependence of $\chi^2$ on the parameters $f_{k,l}$. To this end we define $k(m)$ and $l(m)$ as the indices of the grid square that contains the $m$th measurement, i.e.
\be
(q^{(x)}_m,q^{(y)}_m) \in \{k(m),l(m)\}
\ee
and define $\xi_m$ and $\eta_m$ as the value of the dimensionless coordinate on that grid square corresponding to $q^{(x)}_m$ and $q^{(y)}_m$ (just as in~(\ref{eq:t2})):
\be
\xi_m \equiv \frac{q^{(x)}_m-x_{k(m)}}{w_{k(m)}^{(x)}},\quad\quad \eta_m \equiv \frac{q^{(y)}_m-y_{l(m)}}{w_{l(m)}^{(y)}}
\ee
Also, in order to be able to express $\partial S/\partial x$ and $\partial S/\partial y$ let us define the following matrices:
\be
\begin{split}
E_{m}^{n_1,n_2} &= \sum\limits_{i,j=0}^{3} (Y^{-1})_{4l(m)+j}^{n_2} (X^{-1})_{4k(m)+i}^{n_1} \cdot i\xi_m^{i-1} / w^{(x)}_{k(m)} \cdot \eta_m^j\\
F_{m}^{n_1,n_2} &= \sum\limits_{i,j=0}^{3} (Y^{-1})_{4l(m)+j}^{n_2} (X^{-1})_{4k(m)+i}^{n_1} \cdot \xi_m^i \cdot j\eta_m^{j-1} / w^{(y)}_{l(m)}
\end{split}
\ee
With the matrices $E$ and $F$ the expression for $\chi^2$ in~(\ref{eq:chisqr}) can be rewritten as:
\be
\chi^2(f_{k,l})= \sum\limits_{m=0}^{N-1}\left[ \left( \frac{E_m^{n_1,n_2}f_{n_1,n_2} - D^{(x)}_m}{\Delta D^{(x)}_m}\right)^2 + \left( \frac{F_m^{n_1,n_2}f_{n_1,n_2} - D^{(y)}_m}{\Delta D^{(y)}_m} \right)^2 \right]
\ee
which implies that in the system of linear equations~(\ref{eq:sys}) to be solved appear
\begin{align}
M_{n_1,n_2}^{k,l} &= \sum\limits_{m=0}^{N-1} \left [ \left(\Delta D^{(x)}_m\right)^{-2} E_m^{k,l}E_m^{n_1,n_2} + \left(\Delta D^{(y)}_m\right)^{-2} F_m^{k,l}F_m^{n_1,n_2} \right]\\
V_{n_1,n_2} &= \sum\limits_{m=0}^{N-1} \left [ \left(\Delta D^{(x)}_m\right)^{-2} E_m^{n_1,n_2} D^{(x)}_m + \left(\Delta D^{(y)}_m\right)^{-2} F_m^{n_1,n_2} D^{(y)}_m \right]
\end{align}
Using the actual form of $M$ and $V$ the system of linear equations in~(\ref{eq:sys}) can be solved\footnote{The system of linear equations can be solved using e.g. the Lapack library.} for $f_{k,l}$. With the spline parameters the spline coefficients are also determined through~(\ref{eq:c2}). 

Since the number of measurements is $2N$ and we have $K\cdot L-1$ independent spline parameters, the degrees of freedom of this fit is given by the difference $\textmd{dof}=2N-K\cdot L+1$. This way the freedom corresponding to the translational symmetry is transformed out: since $f_{0,0}=0$, the result $S(x,y)$ is now set such that $S(x_0,y_0) = 0$ holds. Note that the spline function can also be transformed easily to satisfy some a priori known reference equation $S(x_r,y_r) = S_r$ by a simple translation $S\to S+(S_r-S(x_r,y_r))$.

We note that a similar approach to fit a spline function to given gradient information was presented in Ref.~\cite{Juttler:2002}.

\section{Stable solutions}
\label{sec:stable}

The method described in the previous section is bound to give the function $S(x,y)$ for which the sum of deviations $\chi^2$ is the smallest. The solution on the other hand also depends on the number and the position of the nodepoints, and these have to be tuned appropriately in order to determine the surface that fits best.

If the number of nodepoints $K\cdot L$ is small compared to the number of measurements $N$, then the reduced sum of deviations will be large ($\chi^2/\textmd{dof} \gg 1$) and the spline function $S(x,y)$ may not be a good approximation to the surface sought for. On the other hand if $K\cdot L$ is large\footnote{Obviously the inequality $K\cdot L-1<2N$ should hold otherwise the problem is underdetermined.}, the best fit will become an oscillatory function which has the correct derivatives everywhere (i.e. $\chi^2/\textmd{dof}\approx 1$), but is probably not the ``right'' solution, especially when one knows that $S$ should be monotonic\footnote{For example this is the case for the pressure $\log\mathcal{Z}$ as a function of the temperature.} in e.g. $x$. This unwanted feature is a characteristic of spline functions even in one dimension. Note that this problem can also be solved using B-splines, where the monotonicity of the surface can be guaranteed by imposing linear inequality constraints.

\begin{figure}[h!]
\centering
\includegraphics*[width=7.0cm]{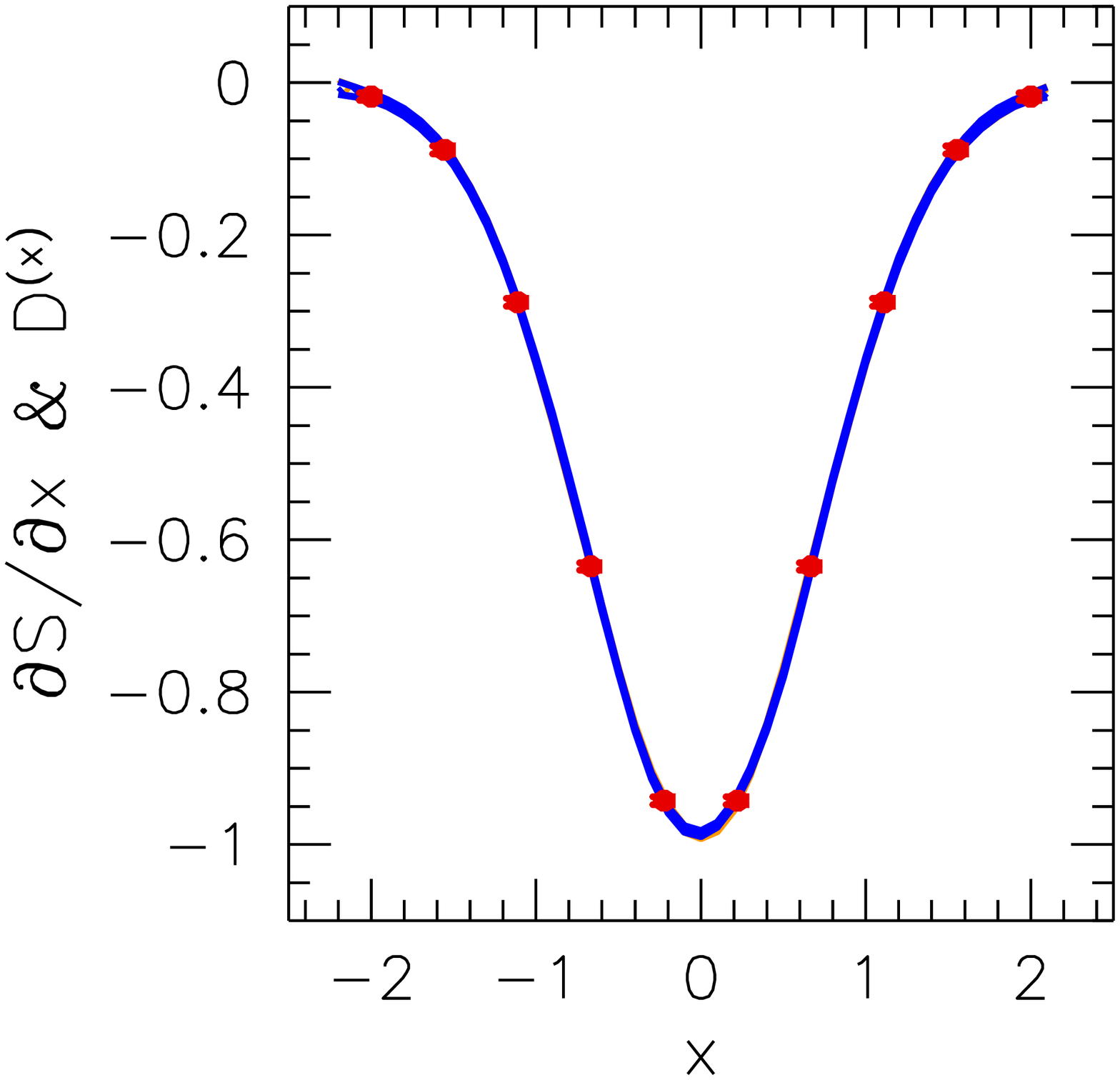} \quad
\includegraphics*[width=7.0cm]{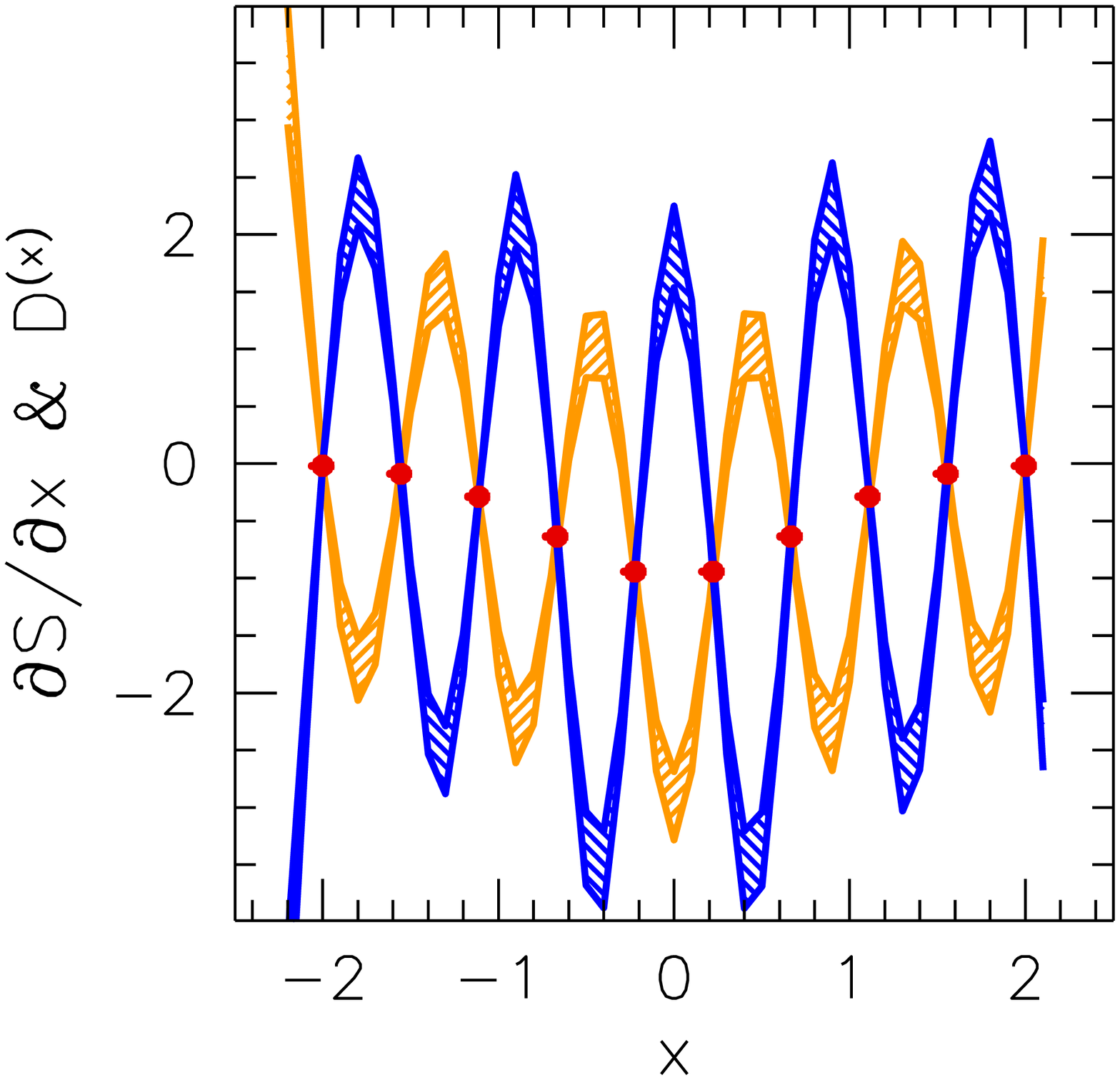}
\caption{One dimensional slice of a two dimensional fit. Shown are the input data $D^{(x)}$ (red points) and the derivative $\partial S/\partial x$ of the spline surface obtained using two nodepoint sets (blue and orange bands). The two sets differ only in one gridpoint, see~(\ref{eq:nodep}). On the left side the number of nodepoints in the $x$ direction is $K=10$; a slight change in the nodepoints results in no visible change in the solution (orange and blue curves are on top of each other). On the other hand, on the right side $K=14$, and a similar change dramatically distorts the solution. The corresponding values for $\mathcal{D}$ are $0.004$ (left) and $\sim10^7$ (right). The width of the band represents the statistical error.}
\label{fig:stable}
\end{figure}

Thus we need a measure of how ``right'' the solution is. A useful way to define this property is to investigate how much $f$ changes as the nodepoints are modified, since the oscillatory solutions are unstable even under a small change in the nodepoints. This way we can filter out the stable, realistic solutions (see illustration on figure~\ref{fig:stable}).

We define modified nodepoint sets $x^{(\alpha)}$ as
\be
\begin{split}
x^{(\alpha)}_k &= x_k+\varepsilon, \quad\quad \textmd{if } k=\alpha \\
x^{(\alpha)}_k &= x_k, \quad\quad\quad\;\,\, \textmd{otherwise}
\end{split}
\label{eq:nodep}
\ee
with $\varepsilon$ some small number, e.g. $\varepsilon=(x_{K-1}-x_0)/K/10$, and in the same manner for $y^{(\beta)}$. Now we carry out the fitting procedure for each of the modified nodepoint sets, resulting in the modified solutions $f^{(\alpha)}$ and $f^{(\beta)}$. The sum of the relative differences between the original solution $f$ and the modified solutions
\be
\mathcal{D} \equiv \frac{1}{K} \sum_{\alpha = 0}^{K-1} \frac{1}{K\cdot L} \sum\limits_{k,l} \left|\frac{f^{(\alpha)}_{k,l} - f_{k,l}}{f_{k,l}}\right| + \frac{1}{L} \sum_{\beta = 0}^{L-1} \frac{1}{K\cdot L} \sum\limits_{k,l} \left|\frac{f^{(\beta)}_{k,l} - f_{k,l}}{f_{k,l}}\right|
\label{eq:D}
\ee
will indeed serve as an indicator of the stability of the fit. If this relative change $\mathcal D$ is under a few percents then the fit can be considered stable. 

\section{Systematics of the method}

The statistical error $\sigma_{stat}$ of the result from the spline fitting method described above can be determined using the standard jackknife algorithm, i.e. the system of linear equations~(\ref{eq:sys}) needs to be solved for each jackknife sample\footnote{Note that this can be performed in a single Lapack call that solves the equations for different vectors on the right hand side.}. The systematical error on the other hand can be determined by varying the nodepoints $\{x_k,y_l\}$. Based on experience the number of nodepoints may range from $M/2$ to $M$; usually an equidistant nodepoint-set can already produce a small value for $\chi^2/\textmd{dof}$, but increasing the density of nodepoints in areas where the function changes rapidly (i.e. where the measured derivatives are large) can further help to improve the fit quality.

Accordingly, a straightforward way to determine the systematical error is to generate various nodepoint sets with different number (and position) of gridpoints.
Then, for each set $\tau$ of the nodepoints the fit is carried out resulting in a spline function $S_\tau$ and an indicator $G_\tau = \left(\chi^2_{corr}/\textmd{dof}\right)^{-1}$ of the fit quality. Results which are in the above detailed sense not stable should be filtered out at this point.
Then at each point the systematical error $\sigma_{sys}$ is determined by
\be
\sigma_{sys} (x,y) = \sqrt{ \left\langle S_{\tau}(x,y)^2 \right \rangle_G - \left\langle S_{\tau}(x,y) \right \rangle_G^2 }
\ee
with
\be
\left\langle \mathcal{O}_{\tau} \right \rangle_G = \sum\limits_\tau \mathcal{O}_\tau G_\tau \Big/ \sum\limits_\tau G_\tau
\ee
Thus the total error can be estimated to be
\be
\sigma_{tot}=\sqrt{\sigma_{sys}^2+\sigma_{stat}^2}
\ee
\section{Testing the method against mock data}
\label{sec:test}

We tested the spline fitting method in two dimensions against three sets of mock data. Input to the method are the coordinates $q^{(x)}_m$ and $q^{(y)}_m$ and the derivatives $D^{(x)}_m$, $D^{(y)}_m$ with $m=0\ldots N-1$ together with 10-10 jackknife samples at each $m$. The derivatives were generated using an original function $F(x,y)$ and were scattered for the jackknife samples to have normal distribution with a relative width of $\Delta$. In table~\ref{tab}. we tabulate information about the data: the function $F$, the number of measurements, the relative error and the type of the input distribution (aligned on a rectangular grid or randomly distributed). The original function was chosen such that it resembles the example mentioned earlier: $\log\mathcal{Z}$ as a function of the temperature and some other parameter near a crossover transition (here the role of the temperature is played by the variable $x$).

\begin{table}[h!]
\centering
\begin{tabular}{c | c | c | c | c }
\textmd{fit} & $F(x,y)$ & $N$ & $\Delta$ & input \\ \hline \hline
1.&$(y+10)\cdot(2+\tanh(4(x-4)))(2x+3))$ & 400 & $2\%$ & rectangular\\
2.&$(4y^2+2y+3)\cdot(1.5+\tanh(4(x-4)))(6x+3)$ & 1600 & $7\%$ & rectangular \\
3.&$(2.6y^2+2.9y+5)\cdot(4+\tanh(3(x-5)))(3x+2)$ & 400 & $2\%$ & random \\
\end{tabular}
\caption{Summary of the mock examples.}
\label{tab}
\end{table}

After the solution was determined, the fitted surface $S(x,y)$ was compared to the original function and to indicate the agreement a weighted sum of deviations
\be
\begin{split}
\beta^S_m &=  \frac{S(q^{(x)}_m,q^{(y)}_m)-F(q^{(x)}_m,q^{(y)}_m)}{\sigma_{tot}(q^{(x)}_m,q^{(y)}_m)} \\
\overline\beta_S &= \frac{1}{N} \sum\limits_{m=0}^{N-1} \left(\beta^S_m\right)^2
\label{eq:indic}
\end{split}
\ee
was calculated.

In table~\ref{tab2} we show information about the fits: the minimal value for $\chi^2/\textmd{dof}$ and the value of the above constructed indicator $\overline\beta_S$. In order to test the method and have a comparison, we also carried out a usual two dimensional integration for the rectangular inputs.
This was done by integrating a one dimensional spline of the $x$-derivatives along the horizontal $x$-direction upto $q^{(x)}_m$, then the $y$-derivatives along the vertical $y$-direction upto $q^{(y)}_m$. Then we repeated this procedure in the opposite order, and the difference in the results was used to define the systematical error for this method. This way, in the same manner as in~(\ref{eq:indic}) we also defined (for rectangular input data) the indicator $\overline\beta_I$, which is also shown in the table. In order to compare the two methods it is also instructive to study the relative statistical $\delta_{stat}$ and relative systematical error $\delta_{sys}$ of the results. In the table we show the average of these quantities $\overline \delta_{stat}$ and $\overline \delta_{sys}$ for both procedures.

\begin{table}[h!]
\centering
\begin{tabular}{c | c | c | c | c | c | c | c}
\textmd{fit} & $\chi^2/\textmd{dof}_{min}$ & $\overline \delta^S_{stat}$ & $\overline\delta^S_{sys}$ & $\overline\beta_S$ & $\overline\delta^I_{stat}$ & $\overline\delta^I_{sys}$ & $\overline\beta_I$ \\ \hline \hline
1. & 1.19 & $0.14\%$ & $0.27\%$ & $0.47$ & $0.52\%$ & $0.82\%$ & $0.64$\\
2. & 1.07 & $0.37\%$ & $0.09\%$ & $0.74$ & $1.66\%$ & $1.36\%$ & $0.35$\\
3. & 1.33 & $0.25\%$ & $0.44\%$ & $0.41$ & - & - & - \\
\end{tabular}
\caption{Fit information for the mock examples.}
\label{tab2}
\end{table}

\begin{figure}[h!]
\centering
\includegraphics*[width=7.0cm]{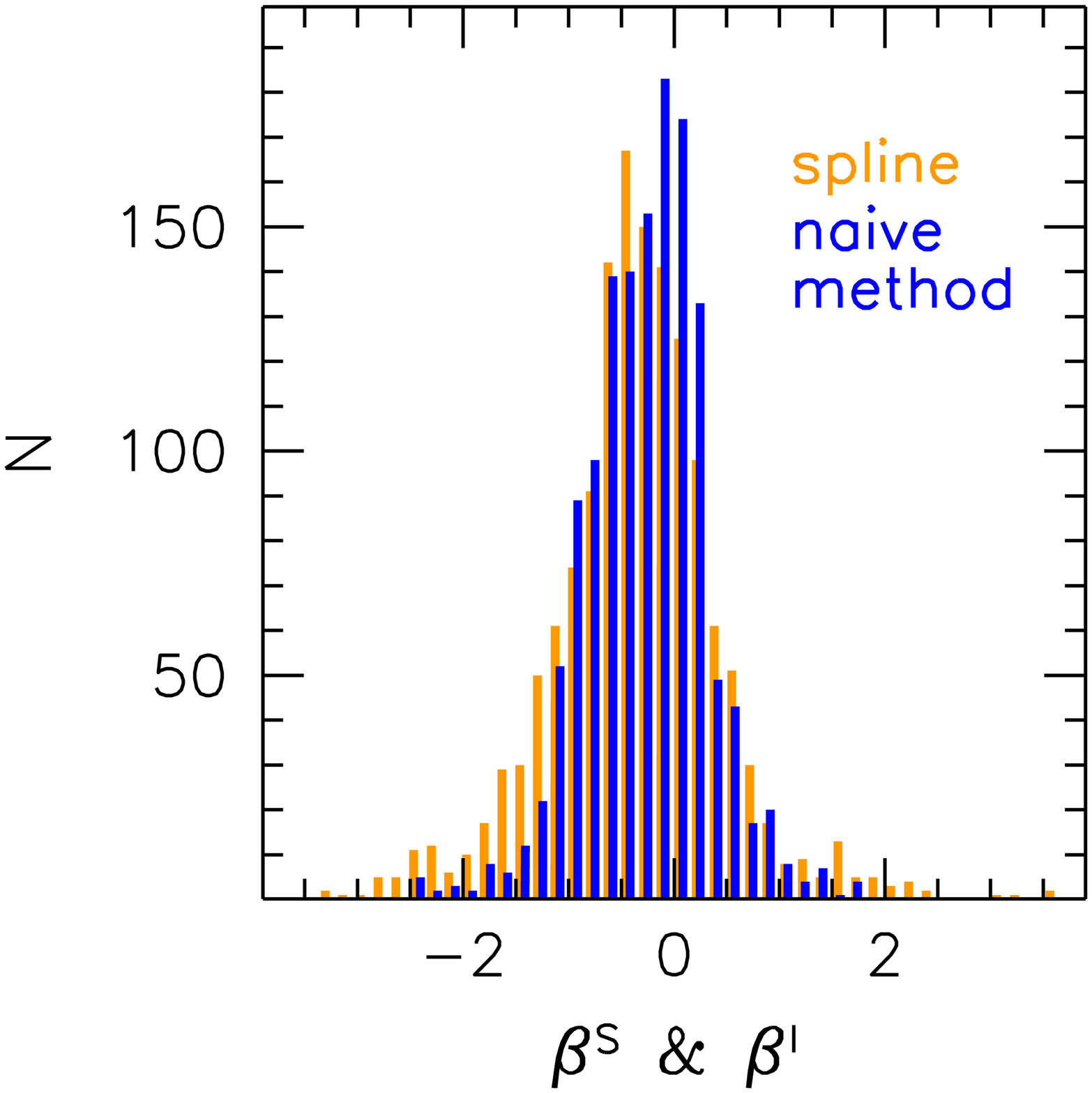}
\includegraphics*[width=7.0cm]{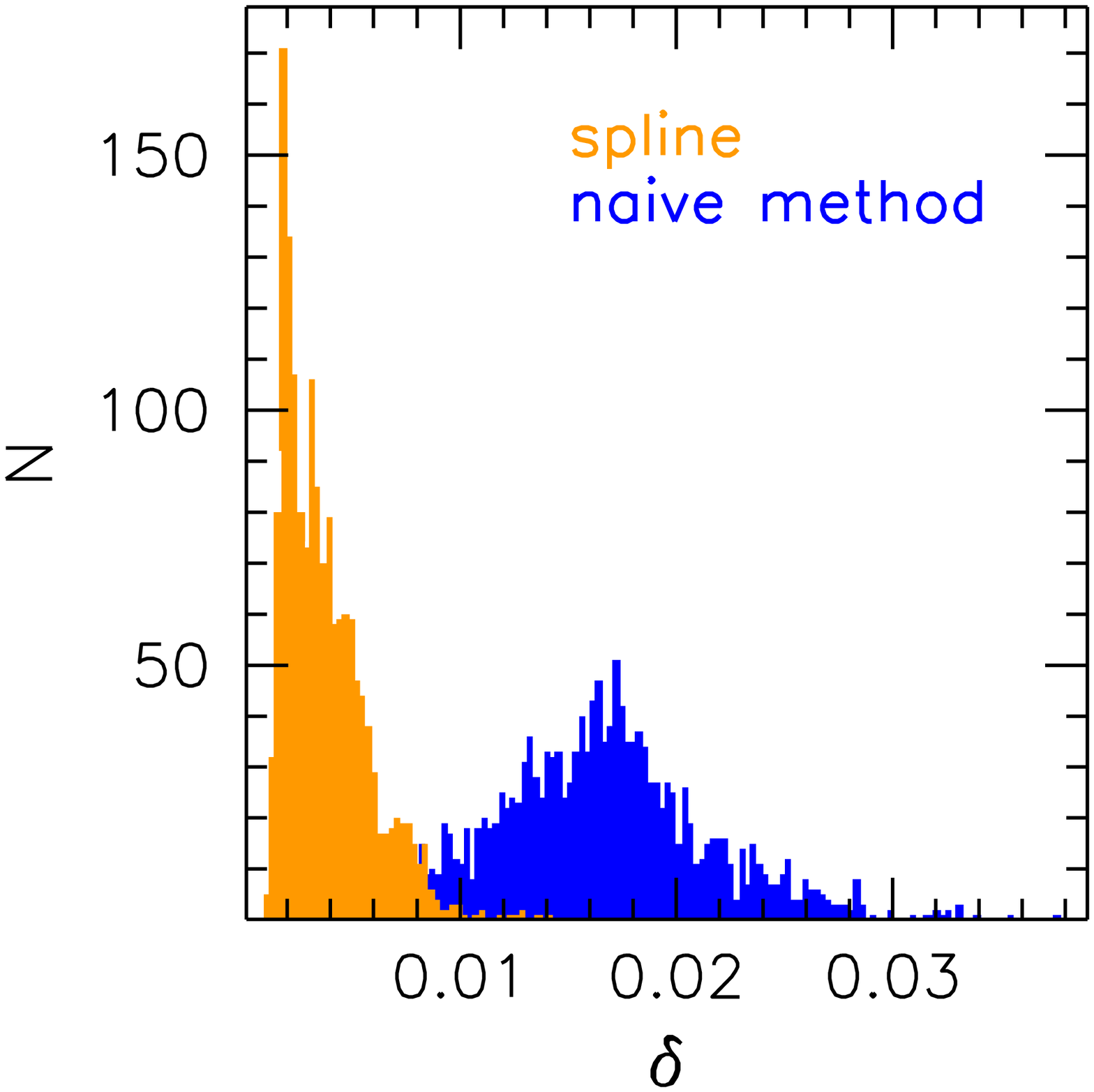}
\caption{Comparison of the results for fit $\#2$. Histogram of the relative deviation $\overline\beta_S$ and $\overline\beta_I$ between the fitted surface and the original function (left side) and histogram of the relative error $\delta = \sqrt{\delta_{stat}^2+\delta_{sys}^2}$ of the result (right side). We show the results using the spline method with orange, and results obtained by the usual integration with blue. Note that while both procedures can be considered as fits with good quality (since the distribution of the deviation $\beta$ is narrow), the relative error is much smaller for the spline integration scheme.}
\label{fig:comp}
\end{figure}

In all cases results obtained by both methods are consistent with each other within errors. A small value for $\overline\beta$ also indicates that the results are indeed good approximations of the original function $F(x,y)$. To confirm this we also show in the left side of figure~\ref{fig:comp} a histogram of the deviation $\beta^S_m$ and $\beta^I_m$ for fit $\#2$, which are well approximated by normal distributions of width $\lesssim 1$.

One should however note that the multidimensional spline fitting algorithm results in much smaller statistical errors than the usual integral method. This is of course due to the fact that in the former case all of the measurements ($\sim N$) are taken into account, while for the latter only those lying on the horizontal-vertical integration path ($\sim 2\sqrt{N}$). The relative error should thus be $\sqrt[4]{N}/\sqrt{2}$ times smaller for the spline fitting method. The fact that this is indeed realized (see table~\ref{tab2}) shows that the method effectively processes the input data. It also is useful to study the distribution of the errors. In the right side of figure~\ref{fig:comp} a histogram for the total relative error $\delta= \sqrt{\delta_{stat}^2+\delta_{sys}^2}$ is plotted for the case of fit $\#2$. This indicates that for the multidimensional spline method errors can be an order of magnitude smaller as compared to the naive integration procedure.

\begin{figure}[h!]
\centering
\includegraphics*[width=5.0cm,angle=-90]{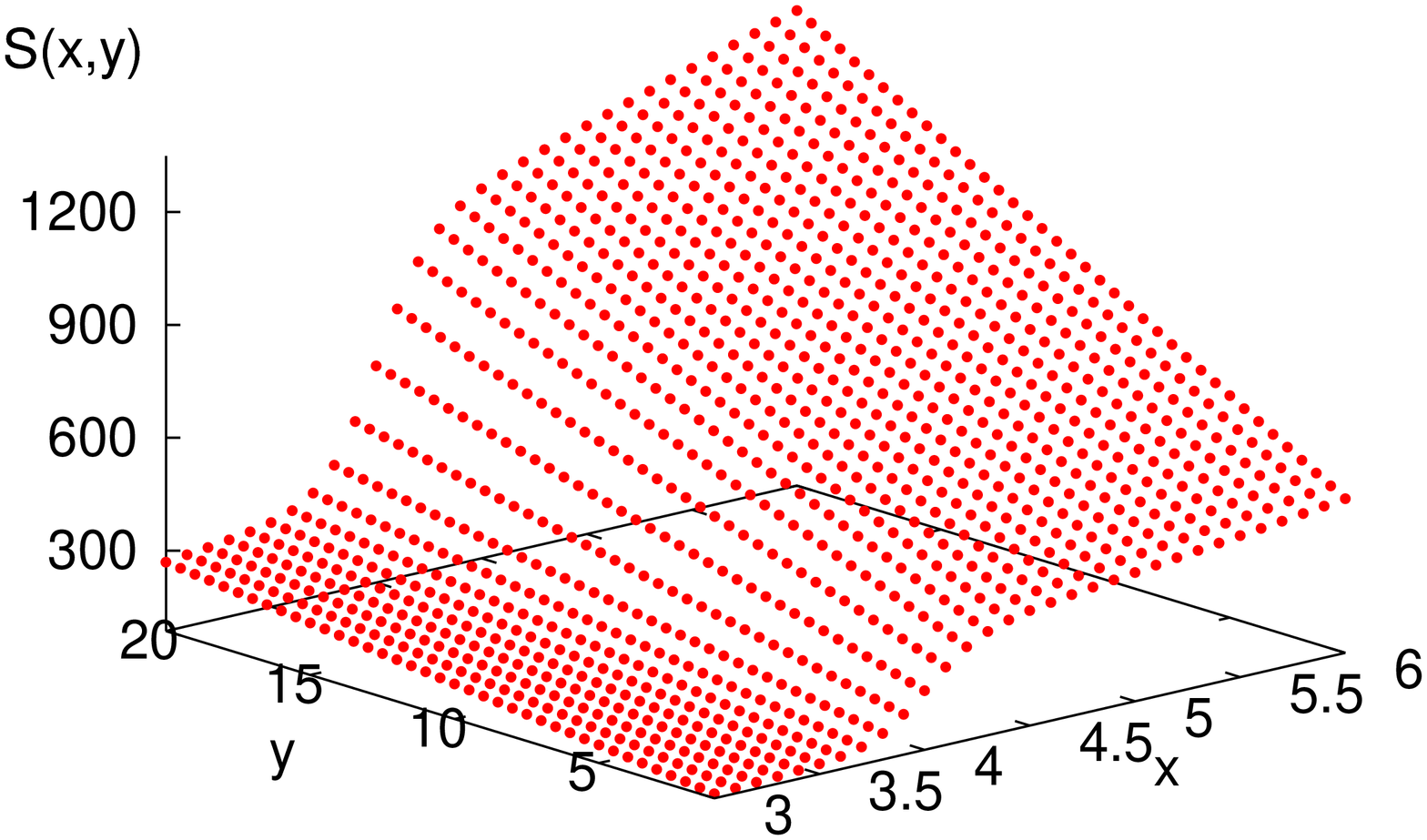}
\includegraphics*[width=5.0cm,angle=-90]{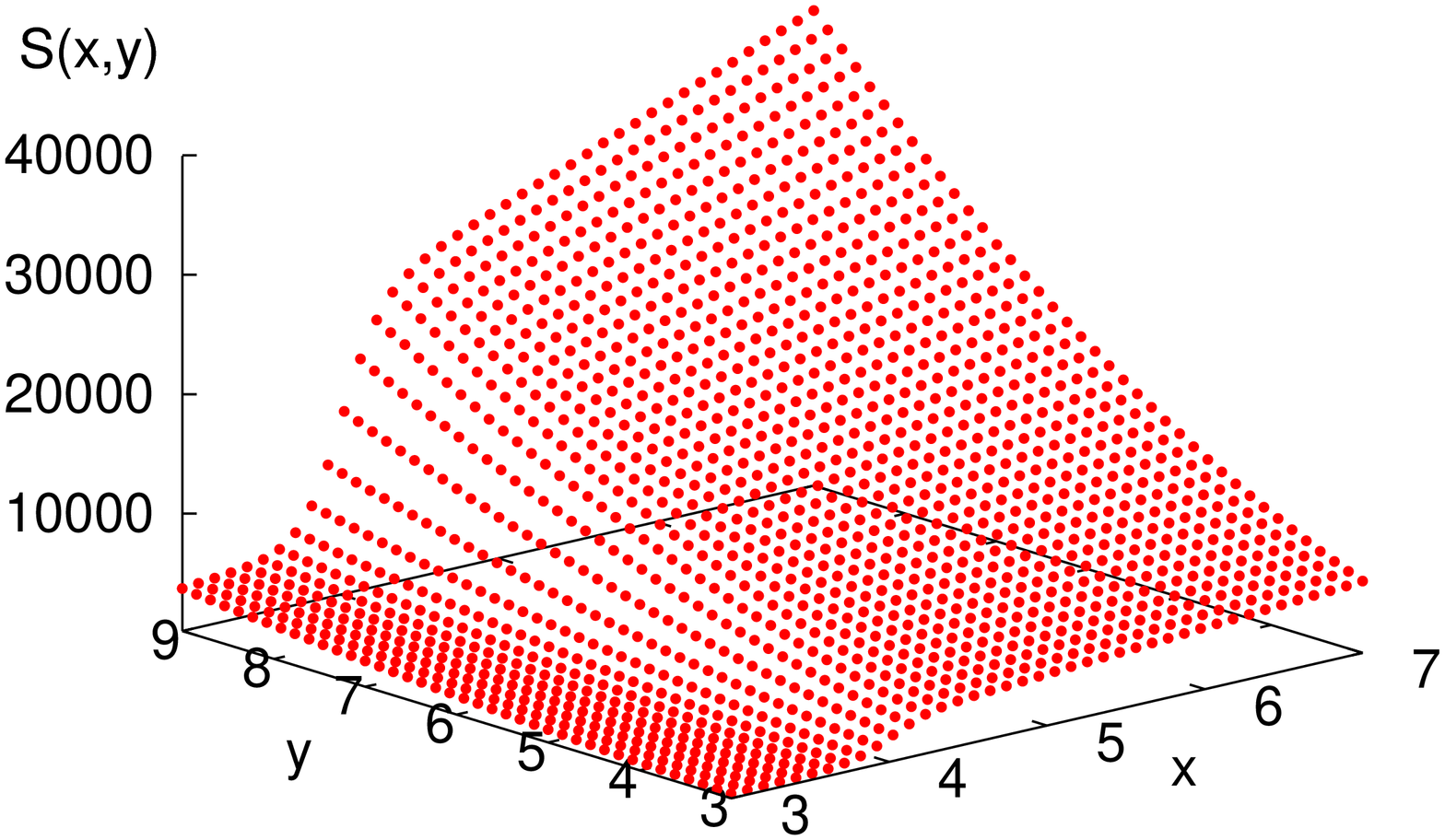}
\includegraphics*[width=5.0cm,angle=-90]{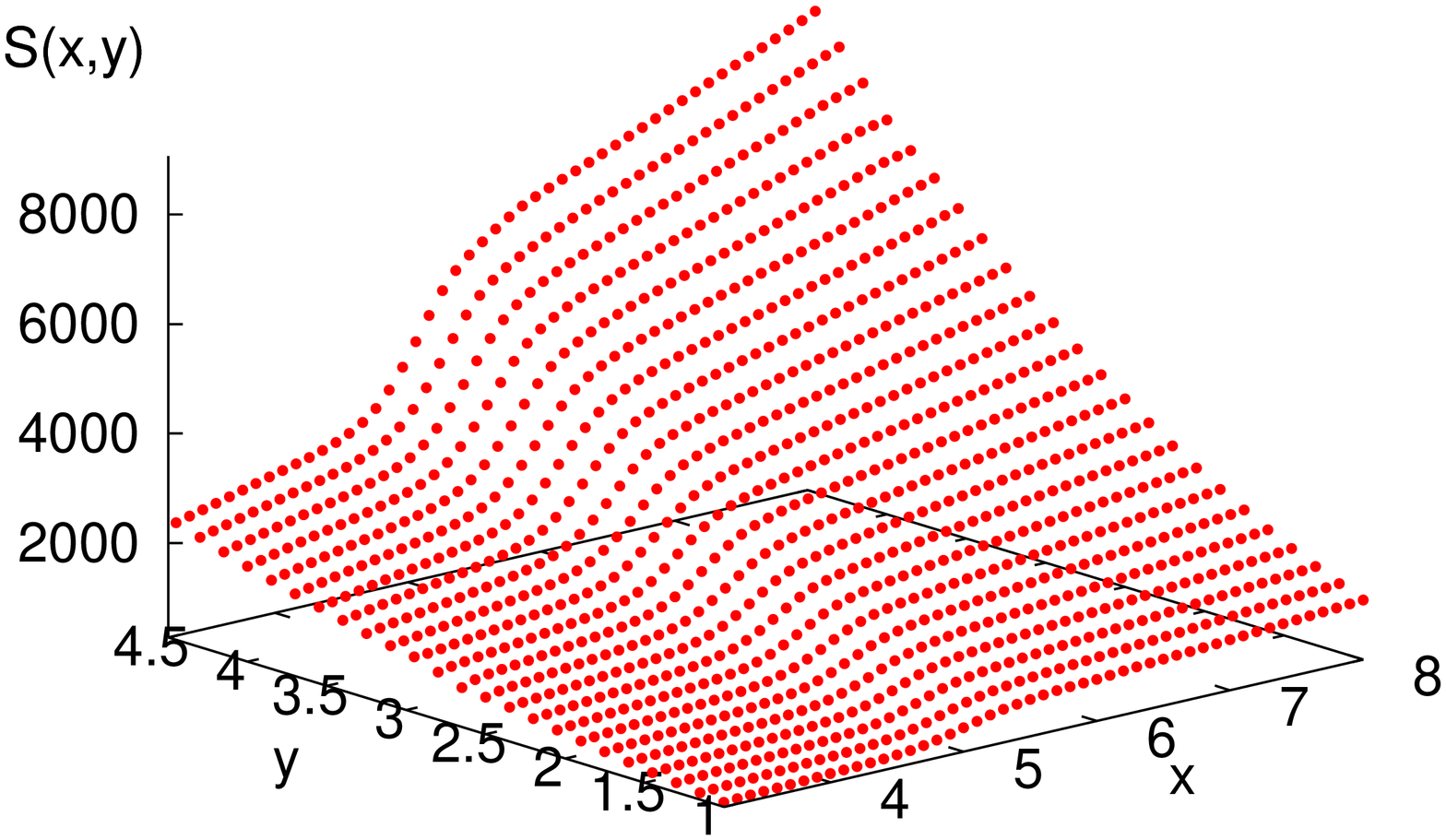}
\caption{The result of the multidimensional spline fitting method for example $\#1$ (left side up), example $\#2$ (right side up) and example $\#3$ (down) The original functions $F(x,y)$ were chosen such that they imitate the behaviour of minus one times the free energy density in the vicinity of a crossover transition as a function of the temperature (here corresponds to $x$) and another parameter like a bare mass.}
\label{fig:3d}
\end{figure}

We remark that for fit $\#1$ a smaller relative error $\Delta$ in the input results in smaller statistical errors, while a large number of input points turns up as a small value for the systematical error. This is also expected, since the statistical error is governed by the difference between each jackknife sample, while the systematical error depends more on the input data density. We also observe that for the case where the input data were randomly distributed, the systematical error dominates over the statistical one. For illustration we plot the results $S(x,y)$ for all three examples on figure~\ref{fig:3d}.

Besides being able to effectively use the increased statistics, the main advantage of the spline fitting procedure is that it better estimates the systematics of the result. While for the two-dimensional integration only two paths are considered (the vertical-horizontal and the horizontal-vertical paths), the spline method is in some sense equivalent to taking into account all possible integration paths at the same time.
Our results indicate that the contribution coming from these generalized paths cannot be neglected in order to estimate the systematics, and the systematic error of the spline method is significantly reduced as compared to the naive two-dimensional integration. 

\section{Summary}

In this paper we presented an algorithm that determines a smooth hypersurface $S(\mathbf{q})$ using the measured gradient of the surface. The algorithm can thus be applied as a multidimensional integration method, which is often useful if one is interested in a continuous approximation of a function of more than one variables. We demonstrated the method on several examples that resemble a situation of the pressure near an analytic, crossover-like phase transition, as a function of two parameters. The method, however can easily be implemented for an arbitrary number of dimensions. Our examples show that usual integration of the gradients along a single path is ineffective and does not give a good estimate on the systematics. The proposed method on the other hand is capable of taking into account every possible integration path and thus has the advantage that it both decreases the statistical error and better estimates the systematical error of the result. The method is not restricted to input data that resides on a rectangular grid, but is capable of handling the situation when the gradient of the function is measured at scattered values of the parameters $\mathbf{q}$. 

The new method is primarily applicable in statistical physics, e.g. in lattice field theory. Here it constitutes an essential improvement as compared to the typically used methods like the conventional integration. This scope of application also motivated the example surfaces of section~\ref{sec:test}, which resemble continuous phase transitions in the space of some bare parameters. In particular, the present method was utilized to obtain results for the pressure $\log \mathcal{Z}$ using the lattice determined derivatives of $\log \mathcal{Z}$ in Ref.~\cite{Borsanyi:2010cj}.

\section*{Acknowledgements}

This work is supported by the grant (FP7/2007-2013)/ERC no. 208740. The author owes many thanks to S\'andor Katz for stimulating and useful discussions. Furthermore the author thanks S\'andor
Katz, Zolt\'an Fodor and K\'alm\'an Szab\'o for a careful reading of the manuscript.

\newpage

\appendix

\section{The spline matrix}
\label{sec:app1}
\noindent
In the following we list the code that generates the matrix $X$:
\begin{verbatim}

/* Input */
int K;      // Number of grid points in the given direction}
int n_exp;  // Number of exponents; this works for n_exp=4 or 5
double *W;  // Array of size (K-1) containing the grid widths

/* Output */
double *X   // Array of size (n_exp*(K-1))*(n_exp*(K-1)), result

/* Auxiliary variables */
int k,l,p;
int eqno = 0;
int fac[4][5] = {{1,1,1,1,1},{0,1,2,3,4},{0,0,2,6,12},{0,0,0,6,24}};
int n = n_exp * (K-1);

/* fill X with zeros */ 
for (k=0; k<n; k++) for (l=0; l<n; l++) {
    X[k*n+l] = 0.0;
}

/* fix spline value at K points */
for (l=0; l<K-1; l++) {
    X[eqno*n + n_exp*l] = 1.0;
    eqno++;
}	
for (k=0; k<n_exp; k++) X[eqno*n + n_exp*(K-2)+k] = 1.0;
eqno++;

/* fix continuity of zeroth, first, second ... (n_exp-2)-th derivative of spline */
for (l=0; l<K-2; l++) for (p=0; p<n_exp-1; p++) {
    for (k=0; k<n_exp; k++) X[eqno*n + n_exp*l + k] = fac[p][k] / pow(W[l],p); 
    X[eqno*n + n_exp*(l+1) + p] = -fac[p][p] / pow(W[l+1],p);
    eqno++;
}

/* (n_exp-2)-th derivative goes to zero at the ends */
X[eqno*n + n_exp-2] = fac[n_exp-2][n_exp-2] / pow(W[0],n_exp-2);
eqno++;
for (k=0; k<n_exp; k++) 
    X[eqno*n + (K-2)*n_exp + k] = fac[n_exp-2][k] / pow(W[K-2],n_exp-2);
eqno++;

/* for n_exp=5 set second derivative to zero at left end */
if (n_exp == 5) {
    X[eqno*n + n_exp-3] = fac[n_exp-3][n_exp-3] / pow(W[0],n_exp-3);
    eqno++;
}

\end{verbatim}

Using the input values \verb!K=5!, \verb!n_exp=4! and the vector of widths $w$, the output from the above code for $X$ is the following:
\be
\left(
\begin{tabular}{c c c c c c c c c c c c c c c c }
1 & 0 & 0 & 0 & 0 & 0 & 0 & 0 & 0 & 0 & 0 & 0 & 0 & 0 & 0 & 0 \\
0 & 0 & 0 & 0 & 1 & 0 & 0 & 0 & 0 & 0 & 0 & 0 & 0 & 0 & 0 & 0 \\
0 & 0 & 0 & 0 & 0 & 0 & 0 & 0 & 1 & 0 & 0 & 0 & 0 & 0 & 0 & 0 \\
0 & 0 & 0 & 0 & 0 & 0 & 0 & 0 & 0 & 0 & 0 & 0 & 1 & 0 & 0 & 0 \\
0 & 0 & 0 & 0 & 0 & 0 & 0 & 0 & 0 & 0 & 0 & 0 & 1 & 1 & 1 & 1 \\
1 & 1 & 1 & 1 & -1 & 0 & 0 & 0 & 0 & 0 & 0 & 0 & 0 & 0 & 0 & 0 \\
0 & $\frac{1}{w_0}$ & $\frac{2}{w_0}$ & $\frac{3}{w_0}$ & 0 & $\frac{-1}{w_1}$ & 0 & 0 & 0 & 0 & 0 & 0 & 0 & 0 & 0 & 0 \\
0 & 0 & $\frac{2}{w_0^2}$ & $\frac{6}{w_0^2}$ & 0 & 0 & $\frac{-2}{w_1^2}$ & 0 & 0 & 0 & 0 & 0 & 0 & 0 & 0 & 0 \\
0 & 0 & 0 & 0 & 1 & 1 & 1 & 1 & -1 & 0 & 0 & 0 & 0 & 0 & 0 & 0 \\
0 & 0 & 0 & 0 & 0 & $\frac{1}{w_1}$ & $\frac{2}{w_1}$ & $\frac{3}{w_1}$ & 0 & $\frac{-1}{w_2}$ & 0 & 0 & 0 & 0 & 0 & 0 \\
0 & 0 & 0 & 0 & 0 & 0 & $\frac{2}{w_1^2}$ & $\frac{6}{w_1^2}$ & 0 & 0 & $\frac{-2}{w_2^2}$ & 0 & 0 & 0 & 0 & 0 \\
0 & 0 & 0 & 0 & 0 & 0 & 0 & 0 & 1 & 1 & 1 & 1 & -1 & 0 & 0 & 0 \\
0 & 0 & 0 & 0 & 0 & 0 & 0 & 0 & 0 & $\frac{1}{w_2}$ & $\frac{2}{w_2}$ & $\frac{3}{w_2}$ & 0 & $\frac{-1}{w_3}$ & 0 & 0 \\
0 & 0 & 0 & 0 & 0 & 0 & 0 & 0 & 0 & 0 & $\frac{2}{w_2^2}$ & $\frac{6}{w_2^2}$ & 0 & 0 & $\frac{-2}{w_3^2}$ & 0 \\
0 & 0 & 2 & 0 & 0 & 0 & 0 & 0 & 0 & 0 & 0 & 0 & 0 & 0 & 0 & 0 \\
0 & 0 & 0 & 0 & 0 & 0 & 0 & 0 & 0 & 0 & 0 & 0 & 0 & 0 & 2 & 6 \\
\end{tabular}
\right)
\ee

\section{Correlated case}
\label{sec:app2}
If the measured derivatives $D_x$ and $D_y$ are correlated, the $\chi^2$ function contains additional terms. Usually measurements at different $m$ values are independent, but the correlation of $D_x(m)$ and $D_y(m)$ at the same $m$ is not negligible, since e.g. in a lattice calculation these are determined using the same configurations. This leads to the following $\chi^2_{corr}$:
\be
\begin{split}
\chi^2_{corr}=\sum\limits_{m=0}^{M-1}\Bigg[ &{Q_{(m)}^{-1}}_{00}\left(\frac{\partial S}{\partial x} - D_x(m)\right)^2 + \\
&2{Q_{(m)}^{-1}}_{01}\left(\frac{\partial S}{\partial x}-D_x(m)\right)\left(\frac{\partial S}{\partial y}-D_y(m)\right) +{Q_{(m)}^{-1}}_{11}\left(\frac{\partial S}{\partial y} - D_y(m)\right)^2 \Bigg]
\label{eq:chi2}
\end{split}
\ee
where $Q_{(m)}$ is the $2\times2$ correlation matrix consisting of the correlators of the two measured derivatives at the $m$th point. The corresponding system of linear equations has the same form as~(\ref{eq:sys}); only now on the left and right hand side enter
\begin{align}
M_{n_1,n_2}^{k,l} &= \sum\limits_{m=0}^{N-1} \left [{Q_{(m)}^{-1}}_{00} E_m^{k,l}E_m^{n_1,n_2} + 2{Q_{(m)}^{-1}}_{01} (E_m^{k,l}F_m^{n_1,n_2}+E_m^{n_1,n_2}F_m^{k,l})+ {Q_{(m)}^{-1}}_{11} F_m^{k,l}F_m^{n_1,n_2} \right]\\
V_{n_1,n_2} &= \sum\limits_{m=0}^{N-1} \left [ {Q_{(m)}^{-1}}_{00} D_x(m)E_m^{n_1,n_2} + 2{Q_{(m)}^{-1}}_{01} (D_x(m)F_m^{n_1,n_2}+E_m^{n_1,n_2}D_y(m))+ {Q_{(m)}^{-1}}_{11} D_y(m)F_m^{n_1,n_2} \right]
\end{align}

\end{document}